\documentclass{PoS}

\title{
\vspace*{-5cm}
{\normalsize {\rm \hfill{ Edinburgh 2013/21} }} \\
\vspace*{5cm}
A light composite scalar in eight-flavor QCD \\ on the lattice}

\ShortTitle{A light composite scalar in eight-flavor QCD on the lattice}

\author{
Yasumichi Aoki$^a$,
Tatsumi Aoyama$^a$,
Masafumi Kurachi$^a$,
Toshihide Maskawa$^a$,
Kohtaroh Miura$^a$,
Kei-ichi Nagai$^a$,
\speaker{Hiroshi Ohki}$^a$, 
Enrico Rinaldi$^b$,
Akihiro Shibata$^c$,
Koichi Yamawaki$^a$,
and 
Takeshi Yamazaki$^a$ 

\hspace*{55mm} (LatKMI Collaboration) 
\\

$^a$
Kobayashi-Maskawa Institute for the Origin
of Particles and the Universe (KMI), Nagoya University, Nagoya
464-8602, Japan \\
$^b$ 
Higgs Centre for Theoretical Physics, SUPA, School of Physics and Astronomy, 
University of Edinburgh, Edinburgh EH9 3JZ, UK \\
$^c$
Computing Research Center, High Energy Accelerator Research Organization (KEK), 
Tsukuba 305-0801, Japan
\\
E-mail: \email{ohki@kmi.nagoya-u.ac.jp}}

\abstract{
In search for a composite Higgs boson (techni-dilaton) in the walking technicolor,  we
present our preliminary results on the first observation of a light flavor--singlet scalar in a candidate theory for the walking technicolor, 
the $N_f=8$ QCD,
which was found in our previous paper to have spontaneous chiral symmetry breaking together with remnants of the conformality.  
Based on simulations with the HISQ-type action on several lattice sizes with various fermion masses, 
we find evidence of a flavor-singlet scalar meson 
with mass comparable to that of the Nambu-Goldstone pion in both the small fermion-mass region, 
where chiral perturbation theory works, and the intermediate fermion-mass region where the hyperscaling relation holds.
We further discuss its chiral limit extrapolation 
in comparison with other states studied in our previous paper: 
the scalar has a mass much smaller than that of the 
vector meson, which is compared to 
the Nambu-Goldstone pion having a vanishing mass in that limit.

}

\FullConference{31st International Symposium on Lattice Field Theory - LATTICE 2013\\
		July 29 - August 3, 2013\\
		Mainz, Germany}

\begin{document}

\section{Introduction}
Recently a Higgs boson with mass around 125 GeV was discovered at LHC. 
While it 
is consistent with the one in the Standard model (SM), 
there still exists a possibility that it is a composite particle in an underlying strongly coupled gauge theory. 
One such example is 
the techni-dilaton 
predicted as a naturally light fermionic bound state in the walking technicolor 
having approximate scale symmetry and  
a large anomalous dimension $\gamma_m \sim 1$~\cite{Yamawaki:1985zg}. 
A composite Higgs as the techni-dilaton is 
a pseudo Nambu-Goldstone (NG) boson of the spontaneously broken 
approximate scale symmetry and 
is shown to be phenomenologically consistent with
the current LHC data~\cite{Matsuzaki:2012mk}.
Thus the most urgent theoretical task to test the walking technicolor 
would be  to check whether or not such a light flavor-singlet scalar bound state
 exists using first-principle calculations on the lattice. (For reviews on the lattice studies in search 
 for candidates for the walking technicolor see~\cite{plenary}.)

Actually, in a previous work~\cite{Aoki:2013zsa}, 
we observed a flavor-singlet fermionic scalar meson $(\sigma)$ lighter than the "pion" (corresponding to the NG pion 
in the broken phase) in the $N_f=12$ QCD, 
which was 
studied by us in another paper on the same setting~\cite{Aoki:2012eq}
and was consistent 
with a conformal theory. 
Since the conformal theory should have
no bound state ("unparticle") at the exact chiral limit, 
the light bound states are only possible in the presence of the fermion mass $m_f$ 
in such a way that it produces
the confining forces (blowing-up coupling) in the infrared region below the fermion mass scale. 
A light scalar in such theory 
would not be regarded as a composite Higgs boson.
Nevertheless the walking theory should have a similar light scalar bound state 
in a similar conformal dynamics, 
with the role of $m_f$ replaced by the dynamically generated mass of the fermion. 

In this paper we indeed observe 
a light flavor-singlet scalar $\sigma$ in the $N_f=8$ QCD, which was shown to be a good candidate 
for the walking technicolor in our previous work~\cite{Aoki:2013xza}.
The $\sigma$ we observe 
could be a first evidence of a candidate for the composite Higgs as a techni-dilaton on the lattice.
As in $N_f=12$ QCD~\cite{Aoki:2013zsa}, 
we extract the $m_\sigma$ from the correlation function of  the $0^{++}$ fermion bilinear operator, 
which consists of both connected and 
(vacuum subtracted) disconnected contributions.
We find that the $\sigma$ is as light as the NG pion similarly to $N_f=12$ QCD~\cite{Aoki:2013zsa}.
As suggested by our previous work~\cite{Aoki:2013xza} on other quantities, 
an approximate hyperscaling behavior 
is also expected for the $m_\sigma$ in the relatively heavier $m_f$ region, 
while near the chiral limit where the spontaneous chiral breaking effects become dominant, 
the $m_\sigma$ should be described 
by a polynomial function as a perturbation of $m_f$.
We then discuss the chiral limit extrapolation of $m_\sigma$ in a way consistent with the chiral perturbation theory.

In the next section, we explain the simulation setup and the methods for the 
flavor-singlet scalar measurement. 
In Section 3, we show the results on the correlation functions,
the $m_\sigma$ as a function of $m_f$. 
In  Section 4, we summarize our results and discuss 
the implications of light scalar in the chiral limit for a composite Higgs scenario.
All the results shown here are preliminary.

\section{Lattice setup}
The gauge configurations for SU(3) gauge theory with eight fundamental fermions 
are generated by the HMC algorithm with tree-level Symanzik gauge action
and HISQ (highly improved staggered quark) action without 
tadpole improvement and 
mass correction in the Naik term.
By using two degenerate staggered fermion species, 
we carry out simulations on three different lattice volumes $(V=L^3)$ 
$L=18, 24,$ and $30$, with fixed aspect ratio $T/L=4/3$ at a single lattice 
spacing ($\beta \equiv 6/g^2=3.8$) for 
five different fermion masses ($m_f=0.02, 0.04, 0.06, 0.08, 0.10$).
For each parameter, we accumulate more than 5000 trajectories, and 
perform measurements every 2 trajectories.
Such a number of configurations allows us to obtain a reasonable signal of $m_\sigma$. 
The statistical error is estimated by the standard jackknife method with bin size
of more than 100 trajectories.

For the measurement of the flavor-singlet scalar, 
we use the following local staggered fermion bilinear operator 
\begin{equation}
O_S(t) = \sum_i \sum_x \bar{\chi}_i(x,t) \chi_i(x,t), 
  \label{eq:op}
\end{equation}
where $i$ denotes the staggered fermion species, $i=1,2$.
Using this operator we measure the two-point correlation function
$\langle O_S(t) O_S(0)\rangle \propto 2D(t) - C(t)$, 
where $C(t)$ and $D(t)$ are the connected and the vacuum subtracted 
disconnected correlators, respectively. 
The factor $2$ in front of $D(t)$ is due to the number of species.
For the calculation of $D(t)$, which is essential to estimate the $\sigma$ correlator, 
we need to calculate the inverse of the Dirac operator
for all the space-time points $(x,t)$. 
In order to reduce the computational cost of the inversion, 
we use the stochastic estimator with noise vectors for space-time and color.
The large fluctuation coming from random noise can be efficiently reduced 
by employing the noise reduction technique for staggered fermions~\cite{Kilcup:1986dg,Venkataraman:1997xi}, 
which was already applied in previous studies, for example, 
the calculation of the flavor-singlet pseudo-scalar~\cite{Venkataraman:1997xi, Gregory:2007ev},
the chiral condensate~\cite{McNeile:2012xh}, 
and also the $\sigma$ in $N_f=12$ QCD~\cite{Jin:2012dw,Aoki:2013zsa}.
The chosen number of noise vectors for each gauge configuration is $64$
to sufficiently suppress the fluctuation of random noise compared to gauge fluctuation.
We tested the calculation method in the $N_f=12$ QCD case 
as reported in Ref.~\cite{Aoki:2013zsa}. 

In the staggered fermion formulation, 
the scalar operator in Eq.(\ref{eq:op}) overlaps not only with the $\sigma$, 
but also with the pseudo-scalar state $(\pi_{\overline{SC}})$, which
is the staggered parity partner of $\sigma$ and 
has the staggered spin-taste structure $(\gamma_4 \gamma_5 \otimes \xi_4 \xi_5)$.
In order to reduce the contribution from the parity partner 
we use the projection 
$C_+(t) \equiv 2C(t) + C(t+1) + C(t-1)$ at even $t$.
The full correlator $2D_+(t) - C_+(t)$ 
in the large $t$ region behaves as 
\begin{equation}
2D_+(t) - C_+(t) = A_\sigma(t),
 \label{eq:ct}
\end{equation}
where $A_H(t)= A_H(e^{-m_H t} + e^{-m_H(T-t)})$.
The connected correlator $C_+(t)$ 
in the large $t$ region can be regarded as $A_{a_0}(t)$
where $a_0$ is the flavor non-singlet scalar state.
Thus, the asymptotic behavior of $2D_+(t)$ is given by
\begin{equation}
2D_+(t) = A_\sigma(t) + A_{a_0}(t).
  \label{eq:dt}
\end{equation}
This means that both $2D_+(t)-C_+(t)$ and $2D_+(t)$ 
can be used to extract $m_\sigma$ from their ground state masses, 
if $m_\sigma < m_{a_0}$. We will discuss this point later.
Another projection $C_-(t) \equiv 2C(t)- C(t+1)- C(t-1)$ at even $t$
is also used to obtain the $\pi_{SC}$ state, which is 
the parity partner of $a_0$.

\section{Result}
\begin{figure}[htbp]
\centering
   \includegraphics[width=6.5cm,clip]{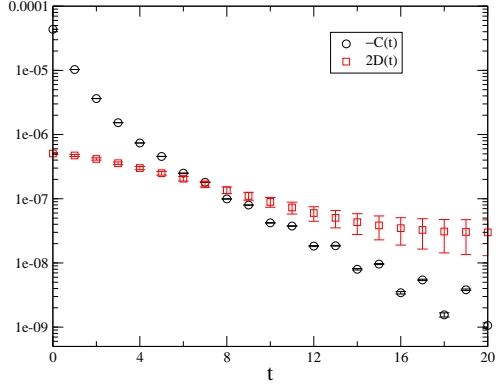} \ \ 
 \caption{
Connected $-C(t)$ and disconnected $2D(t)$ correlators 
for $L=30$, $m_f=0.02$.
}
 \label{fig:corr}
\end{figure}

Figure~\ref{fig:corr} shows a typical result of $-C(t)$ and
$2D(t)$ for $L=30$, $m_f=0.02$. 
As shown in the figure, we can obtain a good signal for 
$2D(t)$ thanks to a large statistics and the noise reduction technique.
In the large $t$ region, $2D(t)$ behaves as a smooth function of $t$.
This result indicates 
that the taste symmetry breaking effects on the parity partner 
are small \cite{Aoki:2013zsa} thanks to utilizing the HISQ-type action.
The smallness of the taste symmetry breaking was also observed in other
meson masses in our previous work~\cite{Aoki:2013xza}.

\begin{figure}[htbp]
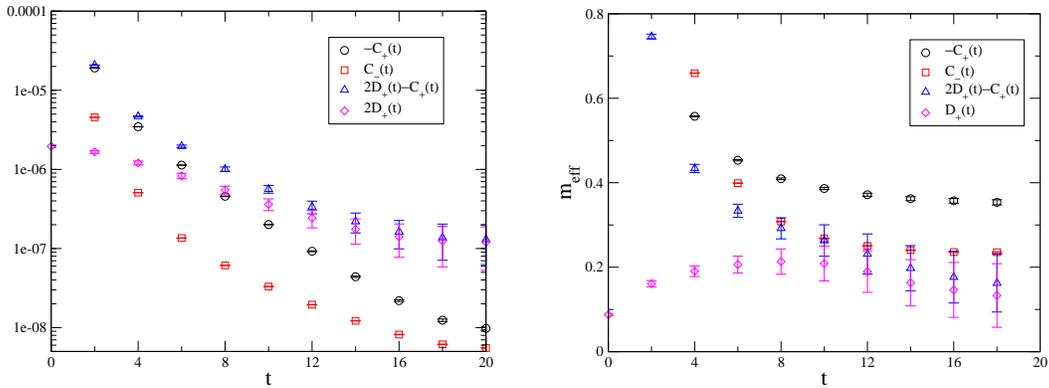

\centering
   \includegraphics[width=6.5cm,clip]{corr_0.02_rep.eps} \quad \quad
   \includegraphics[width=6.5cm,clip]{meff_0.02_rep2.eps}
 \caption{
Parity-projected correlators for the different channels 
constructed from $C(t)$ and $D(t)$ (Left), 
and their effective masses (Right) for $L=30$, $m_f=0.02$.
}
 \label{fig:meff}
\end{figure}

The left panel of Fig.~\ref{fig:meff} shows all the parity-projected correlators 
constructed from 
the $C(t)$ and $D(t)$ for $L=30$, $m_f = 0.02$.
We see that
the full correlator $2D_+(t) - C_+(t)$ at large $t$ 
is dominated by $2D_+(t)$.
Accordingly, the effective mass obtained from  $2D_+(t)$
becomes consistent with that obtained from $2D_+(t) - C_+(t)$
as shown in the right panel of Fig.~\ref{fig:meff}.
This property allows us to evaluate the flavor-singlet scalar mass
from the correlator $2D_+(t)$.
The advantage of using $2D_+(t)$ is that
the plateau appears at small $t$ 
owing to the cancellation 
between the $a_0$ and the contamination from excited states of the $\sigma$.
The plateau of $2D_+(t)$
enables us to determine the effective mass
with relatively smaller statistical error.
We fit $2D_+(t)$ with a single cosh form
in the range $t_{min}=6$ and $t_{max}=T/2$
to obtain $m_\sigma$ for all the values of $m_f$.
Comparing this with the rest of spectrum,
we find that the $\sigma$ is not heavier than 
$\pi_{SC}$,
whose effective mass corresponds to the one of $C_-(t)$ 
in the right panel of Fig.~\ref{fig:meff}.
As for the $a_0$ corresponding to $-C_+(t)$, 
the result would suggest $m_{a_0} > m_\sigma$, as we expected,
although we do not obtain a good effective mass plateau in our lattice volume.

\begin{figure}[tbp]
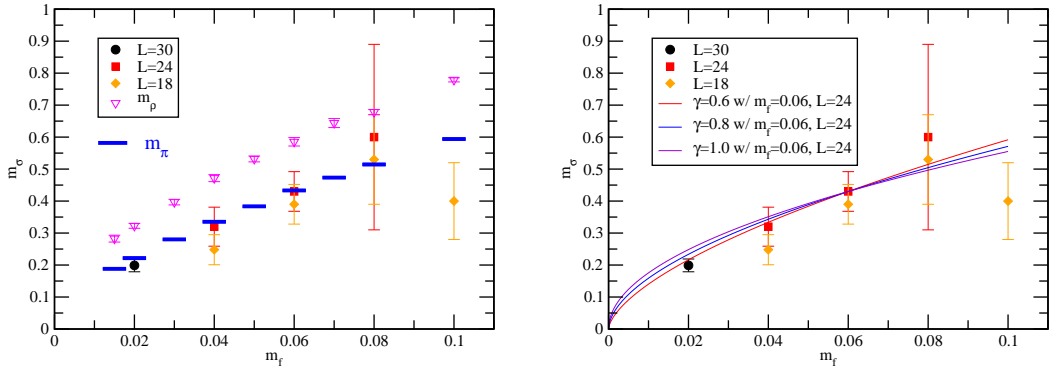

\centering
   \includegraphics[width=6.5cm,clip]{mass_rep8_3.eps} \quad \quad
   \includegraphics[width=6.5cm,clip]{mass_rep4.eps}
 \caption{
(Left) Flavor-singlet scalar mass $m_\sigma$ as a function of $m_f$ for each volume.
The errors are only statistical. For comparison also plotted are $m_\pi$ and $m_\rho$ observed in Ref.~\cite{Aoki:2013xza}.  
(Right) 
Comparison of the $m_\sigma$ data with the hyperscaling curve(s) for the 
value(s) of $\gamma$ observed in Ref.~\cite{Aoki:2013xza} (see text), 
with the absolute value normalized to  the data at $L=24$, $m_f=0.06$.
}
 \label{fig:mass}
\end{figure}

The left panel of Fig.~\ref{fig:mass} presents 
fit results of the $m_\sigma$
as a function of $m_f$ for each volume, together with other masses $m_\pi$ and $m_\rho$ 
taken from Ref.~\cite{Aoki:2013xza}.
The errors are only statistical.
The two results on the different volumes $L=18$ and $24$ at $m_f = 0.04$
are consistent within the statistical error, 
so that
we expect finite size effects to be small at heavier $m_f$.
It is remarkable that all the $m_\sigma$ results at the measured $m_f$ 
are comparable to $m_\pi$. 
This feature is different from that of ordinary QCD with small 
$N_f$,  
and similar to the one observed in $N_f=12$ QCD~\cite{Aoki:2013zsa}. 
It is also to be noted that $m_\sigma$ is much smaller than $m_\rho$.

Since we observed 
an approximate hyperscaling relation of other physical quantities for relatively heavy $m_f$,
$m_f \ge 0.05$, interpreted as a remnant of conformality 
in our previous study~\cite{Aoki:2013xza}, we also expect it for $\sigma$ in this region.
We plot the hyperscaling curves $m_\sigma = C (m_f)^{1/(1+\gamma)}$ 
in the right panel of Fig.~\ref{fig:mass}, where $\gamma = 0.6, 0.8$ and $1.0$ 
are values of 
$m_\pi$, $m_\rho$ and $F_\pi$, respectively, observed  
in Ref.~\cite{Aoki:2013xza}, and 
the value of $C$ is 
matched to the $\sigma$ data at $m_f=0.06$, $L=24$. 
The data are roughly consistent with 
the hyperscaling relation 
with $\gamma=0.6-1.0$ 
within the errors.

On the other hand, the small $m_f$ region, $m_f \le 0.04$,  was shown to be consistent 
with the spontaneous chiral symmetry breaking~\cite{Aoki:2013xza}.
We have seen in  Fig.~\ref{fig:mass} 
that $\sigma$ is much lighter than $\rho$, 
and is as light as $\pi$ all the way down to the
small $m_f$ region well described by the chiral perturbation theory, at least in the present data.  
This would imply a light composite scalar in the chiral limit in the walking theory.

\section{Summary and Discussion}

We have observed for the first time a light flavor-singlet scalar meson 
in the $N_f=8$ QCD, which 
was shown to be a good candidate for the walking technicolor theory 
in our previous study. 
The observed mass is as light as the one of $\pi$ in the simulation parameters region.
Measuring disconnected correlator was critical to the achievement. 
Our results are encouraging in search for the walking technicolor
in view of the 125 GeV Higgs at LHC.

\begin{figure}[tbp]
\centering
   \includegraphics[width=6.5cm,clip]{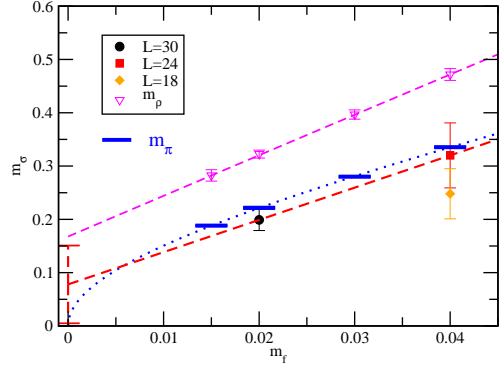}
 \caption{
Fit result of the chiral extrapolation for $m_\sigma$.
For comparison, other spectra of $m_\pi$ and $m_\rho$
and their chiral fits in Ref.~\cite{Aoki:2013xza} are also shown.
}
 \label{fig:mass_fit}
\end{figure}

Although our results are very preliminary and the statistical error is large, 
we discuss the chiral limit extrapolation of $m_\sigma$.
As we observed in the previous paper~\cite{Aoki:2013xza}, 
other physical quantities in the small $m_f$ region are described by the chiral perturbation theory fit.
 In Fig.~\ref{fig:mass_fit} we plot the results of ChPT-like fits of $m_\pi$ and $m_\rho$
with the range $0.015 \leq m_f \leq 0.04$ (4 data)
using the following fit functions~\cite{Aoki:2013xza}, 
\begin{equation}
m_\pi^2 =  c^\pi_1 m_f + c^\pi_2 m_f^2,  \quad \quad
m_\rho =  c^\rho_0 + c^\rho_1 m_f + c^\rho_2 m_f^2.
\label{eq:fit2}
\end{equation}
From the data in this region,
we try to estimate the $m_\sigma$ in the chiral limit. Note that $\sigma$ 
in the chiral limit can be a bound state only in the presence 
of fermion mass dynamically generated by the chiral symmetry breaking, 
which breaks scale symmetry explicitly as well as spontaneously.
Hence its chiral limit mass should not be zero in the same way as $\rho$, 
so that the chiral fit should have the same functional form as that of $\rho$.
We carry out the chiral extrapolation with just the lightest two points on $L \ge 24$ and hence use the linear fit, 
\begin{eqnarray}
m_\sigma &=&  c^\sigma_0 + c^\sigma_1 m_f\, .
\label{eq:fit1}
\end{eqnarray}
The fit result is shown in Fig.~\ref{fig:mass_fit}.
The value in the chiral limit reads $m_\sigma= 0.08(7)$.
Then we have
\begin{eqnarray}
\frac{m_\sigma}{m_\rho}=0.5(5), \quad \quad
\frac{m_\sigma}{F_\pi/\sqrt{2}}=4(4),
\label{eq:chiral}
\end{eqnarray}
in the chiral limit ($F_\pi/\sqrt{2}$ corresponds to $f_\pi=93\, {\rm MeV}$ in the real-life QCD).
Note that in a typical walking technicolor model, 
the one-family (four-weak-doublets) model with $N_f=8$, 
we have $F_\pi/\sqrt{2}\simeq 123\, {\rm GeV}$. 
Within the error 
our results accommodate the 125 GeV Higgs boson.

Here we note another possible signature of the walking behavior 
to be observed on the lattice data.
As we mentioned above, when the chiral symmetry is spontaneously broken, 
the chiral limit of $m_\sigma$ should be non-zero 
due to the very presence of the dynamically generated mass of the fermion, 
while $m_\pi$ should go to zero as a NG boson 
for the same reason (non-zero dynamical mass).
Hence, if $m_\sigma <m_\pi$ for larger $m_f$ as in our data, 
the chiral extrapolation of $m_\sigma$ and $m_\pi$  must be crossing
to $m_\sigma >m_\pi$ at a certain smaller $m_f$.
This never occurs in the conformal phase, since all the masses should obey 
the hyperscaling relations, and the ratio of $m_\sigma/m_\pi$ 
becomes a constant towards the chiral limit. 
Therefore, the observation of  such a crossing phenomenon could be 
another signal of a walking theory. 
This will occur in much smaller $m_f$ region than in the present calculation, 
as seen from our rough chiral extrapolation in Fig.~\ref{fig:mass_fit}.
In order to directly check the crossing on the lattice, we will need 
simulations at even smaller fermion masses and larger volumes.

Besides increasing statistics and obtaining more accurate $m_\sigma$ results in future,
we shall construct a flavor-singlet scalar operator 
by gluonic operators (glueball) to check a consistency between 
the ground state masses extracted from different operators, as 
was studied in the $N_f=12$ QCD~\cite{Aoki:2013zsa, EnricoProceedings}.
We also would need to investigate lattice discretization effects in this theory.

\section*{Acknowledgments}
Numerical calculations have been carried out
on the high-performance computing system $\varphi$ at KMI, Nagoya University, 
and the computer facilities of the Research Institute 
for Information Technology in Kyushu University.	
This work is supported by the JSPS Grant-in-Aid for Scientific Research 
(S) No.22224003, (C) No.23540300 (K.Y.), for Young Scientists (B) No.25800139 (H.O.) and No.25800138 (T.Y.), 
and also by Grants-in-Aid of the Japanese Ministry for Scientific Research on Innovative Areas No.23105708 (T.Y.). 
E.R. was supported by a SUPA Prize Studentship 
and a FY2012 JSPS Postdoctoral Fellowship for Foreign Researchers (short-term).

\end{document}